\documentclass{kluwer}
\newdisplay{guess}{Conjecture}
\begin{document}
\begin{article}
\begin{opening}
\tolerance=10000 \hbadness=10000
\def\verbfont{\small\tt} 
\def\note #1]{{\bf #1]}} 
\def\be{\begin{equation}}
\def\ee{\end{equation}} 
\def\bearr{\begin{eqnarray}} 
\def\eearr{\end{eqnarray}}
\def\barr{\begin{array}}
\title{SEARCH FOR SPATIAL VARIABILITY IN THE SOLAR ACOUSTIC SPECTRUM}
\author{P. \surname{Venkatakrishnan}\thanks{On lien from Indian Institute of Astrophysics, Bangalore, India}}
\author{Brajesh \surname{Kumar}}
\author{S. C. \surname{Tripathy}}
\runningauthor{P. Venkatakrishnan et al.}
\runningtitle{SEARCH FOR SPATIAL VARIABILITY IN THE SOLAR ACOUSTIC SPECTRUM}
\institute{Udaipur Solar Observatory, Physical Research Laboratory, Off Bari Road, 
Dewali, P. B. No. 198, Udaipur 313 001, INDIA 
(E-mails: pvk@uso.ernet.in, brajesh@uso.ernet.in, sushant@uso.ernet.in} 
\date{}
\begin{abstract}
Motivated by the various examples of spatial variability in the power of the
acoustic spectrum, we attempted to look for spatial variability in the peak 
frequency of the spectrum.
However, the determination of this peak frequency on a spatial scale
of a single pixel (8 arc seconds for the GONG data) is limited by the 
stochastic variations in the power spectrum presumably caused 
by the stochastic nature of the excitation process. Averaging over a large 
number of spectra (100 spectra from a 10 $\times$ 10 pixels area) produced
stabler spectra. The peak frequencies of 130 such locations were found to be 
distributed with a FWHM of about 130 $\mu$Hz. A map of the spatial variation of 
this peak frequency did not show any strong feature with statistically 
significant deviation from the mean of the distribution. Likewise, the scatter 
in the peak frequencies masked the detection of magnetic field induced changes 
in the peak frequency. On a much larger scale, the N latitudes showed a
slightly
lower value of the peak frequency as compared to the S latitudes, although the
difference (25 $\mu$Hz) is barely larger than the {\it rms} spread (20 $\mu$Hz).

\end{abstract}

\end{opening}

\section{Introduction} 
The global oscillations of the Sun as measured by Dopplergrams
produced in photospheric lines have maximum power around
5 min (Noyes, 1967, and references therein). The global nature of the 
oscillations was hitherto
exploited by studying the ${\ell-\nu}$ diagnostic diagram, that used
information obtained from each pixel on the solar disk in a
collective manner (Deubener, 1972, and references therein). This technique has 
been extended in recent
times to smaller pieces of the solar disk and to shorter
timescales to produce local helioseismology using a variety of methods such as, 
Fourier-Hankel decomposition (Braun {\em et al.}, 1987), ring diagrams (Hill, 1988), 
acoustic-imaging (Braun {\em et al.}, 1992; Chang {\em et al.}, 1997), 
time-distance seismology (Duvall {\em et al.}, 1993), 
and helioseismic holography (Lindsey and Braun, 1998). Kumar {\em et al.} (2000) 
carried this exercise to the smallest
possible resolution when they demonstrated that the shape and
strength of the acoustic spectrum changed significantly in a
sunspot relative to the quiet Sun. The variation in the shape of the spectrum
can be quantified in terms of the variation in the peak
frequency of the power spectrum. This parameter is relatively free of the effects of 
viewing angle and can therefore be readily adopted for studying the spatial 
variability of the acoustic spectrum. In this
context, the spatial variability of the acoustic power over the solar surface
is well documented, e.g., it has been seen that the absorption efficiency 
of p-mode power increases with the increasing magnetic field density 
(Braun, LaBonte and Duvall, 1990). The question is whether the peak frequency
of the acoustic spectrum could be used as yet another parameter to
map the spatial variability of the acoustic spectrum. In what follows, we
attempt a preliminary investigation of this question using GONG data.

\section{The data analysis}
The data used here are from a time series of Dopplergrams for 11 and 12 May, 1997
for 560 minutes duration starting at 02:30 UT obtained by the
GONG instrument at Udaipur Solar Observatory with a cadence of
1 minute. The Dopplergrams were derotated for mean solar rotation and
registered with the first Dopplergram of each day. It is well known that 
the Dopplergrams, in addition to the {\it p}-modes, also exhibit variety of 
features, such as supergranulation pattern, meridional flows and solar rotation 
gradients. A two-point backward
difference filter (GRASP/IRAF software package),

\begin{equation}
	Filtered Image(t) = Image(t) - Image(t-1)
\end{equation}

was applied to the Dopplergrams to enhance the {\it p}-mode
oscillations above the other features. The temporal power spectrum was
obtained for each pixel. This spectrum was smoothed by applying a digital filter, 
namely a Savitzky-Golay (S-G) filter (Press {\it et al.}, 1992),
to the power spectrum. This filter basically smooths the data by a window 
function of a predefined number of data points and a polynomial order with a proper 
weighting and then finds the maximum of the smoothed spectrum. We observe an
optimal fit with a window of 32 data points and a polynomial of 
order 6. The peak frequency of the smoothed spectrum was then determined by selecting 
the discrete frequency in the appropriate frequency domain whose ordinate was 
greatest in the smoothed spectrum.  

\input epsf
\begin{figure}
\begin{center}
\leavevmode
\epsfxsize=2.75in\epsfbox{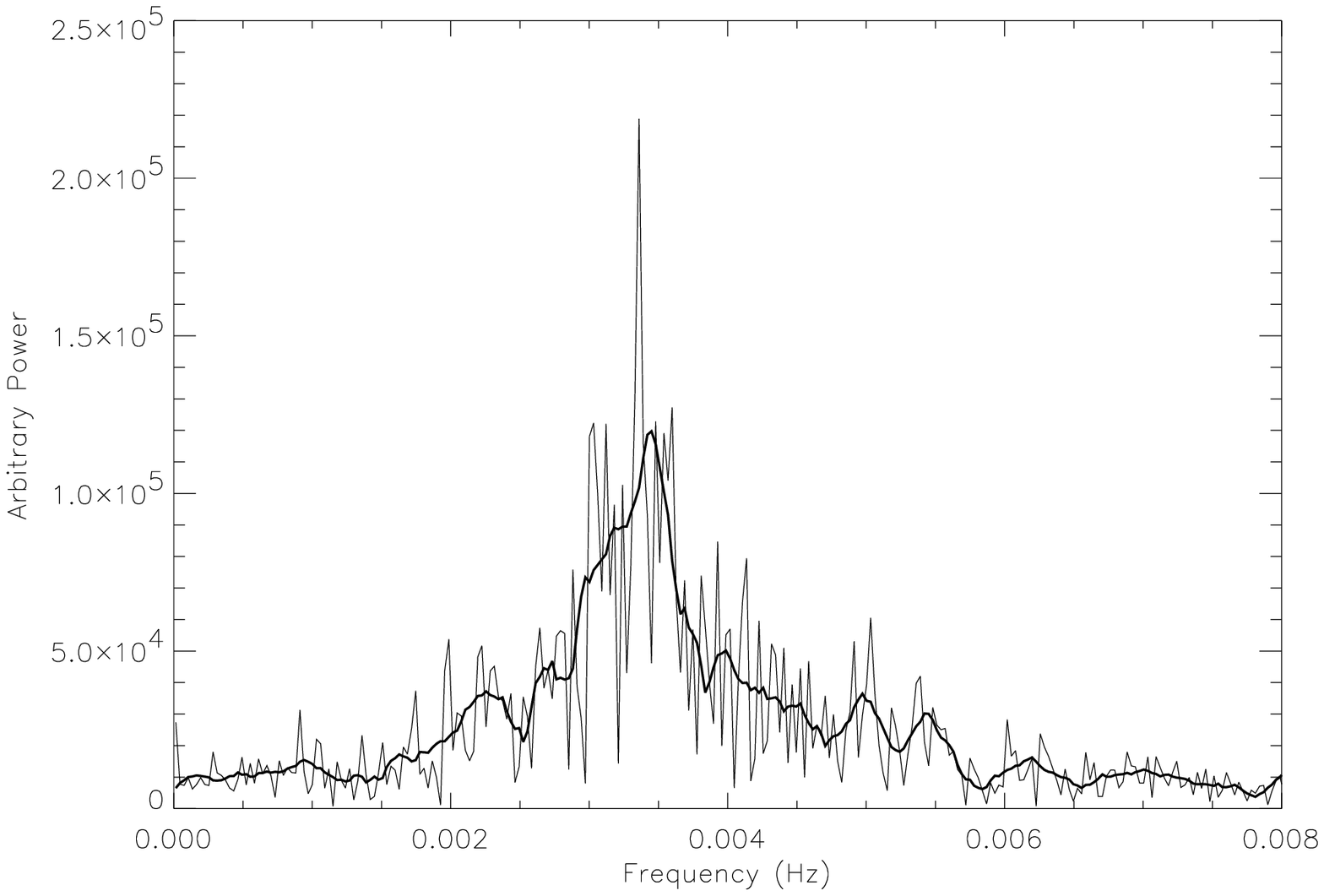}
\epsfxsize=2.75in\epsfbox{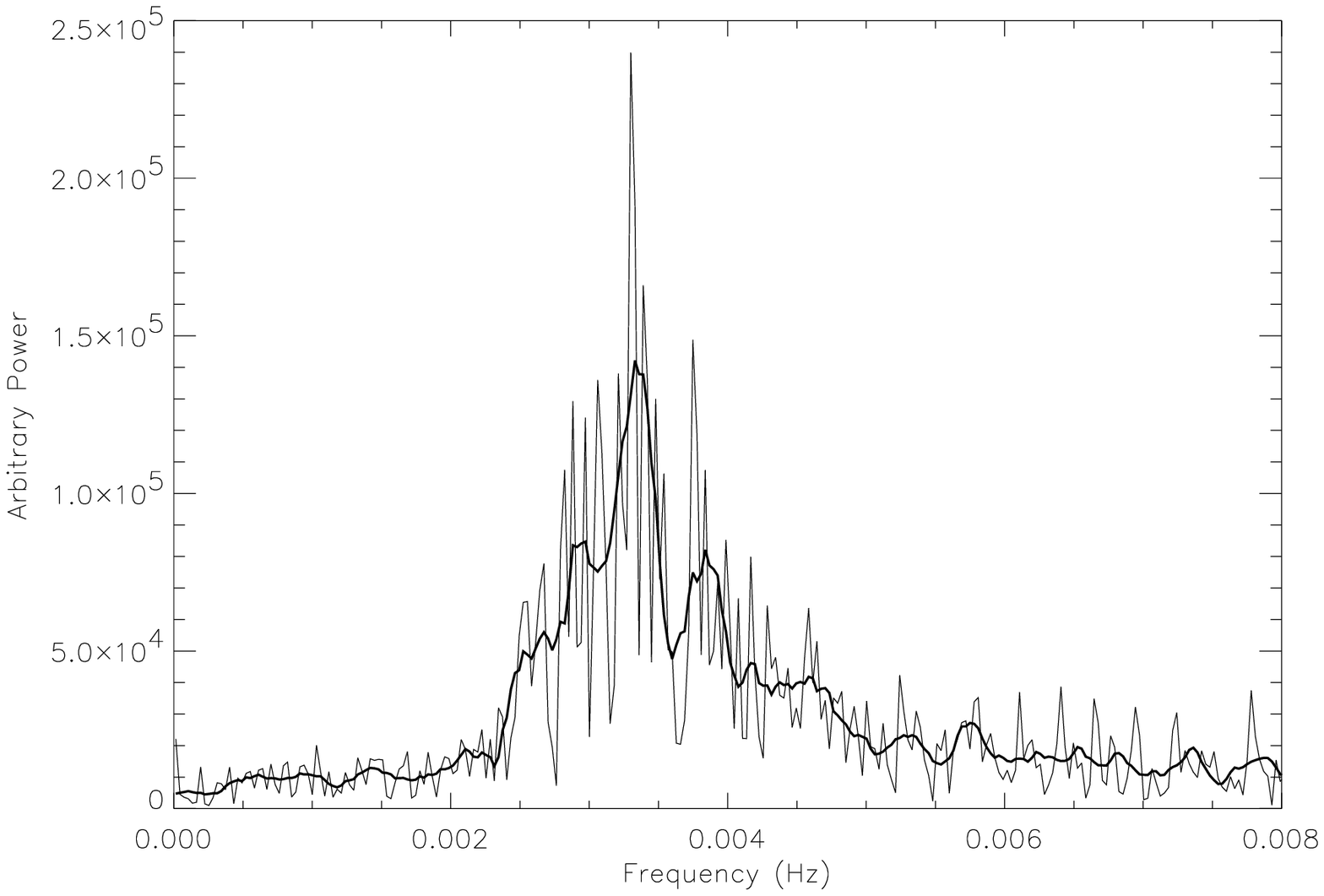}\\
\leavevmode
\epsfxsize=2.75in\epsfbox{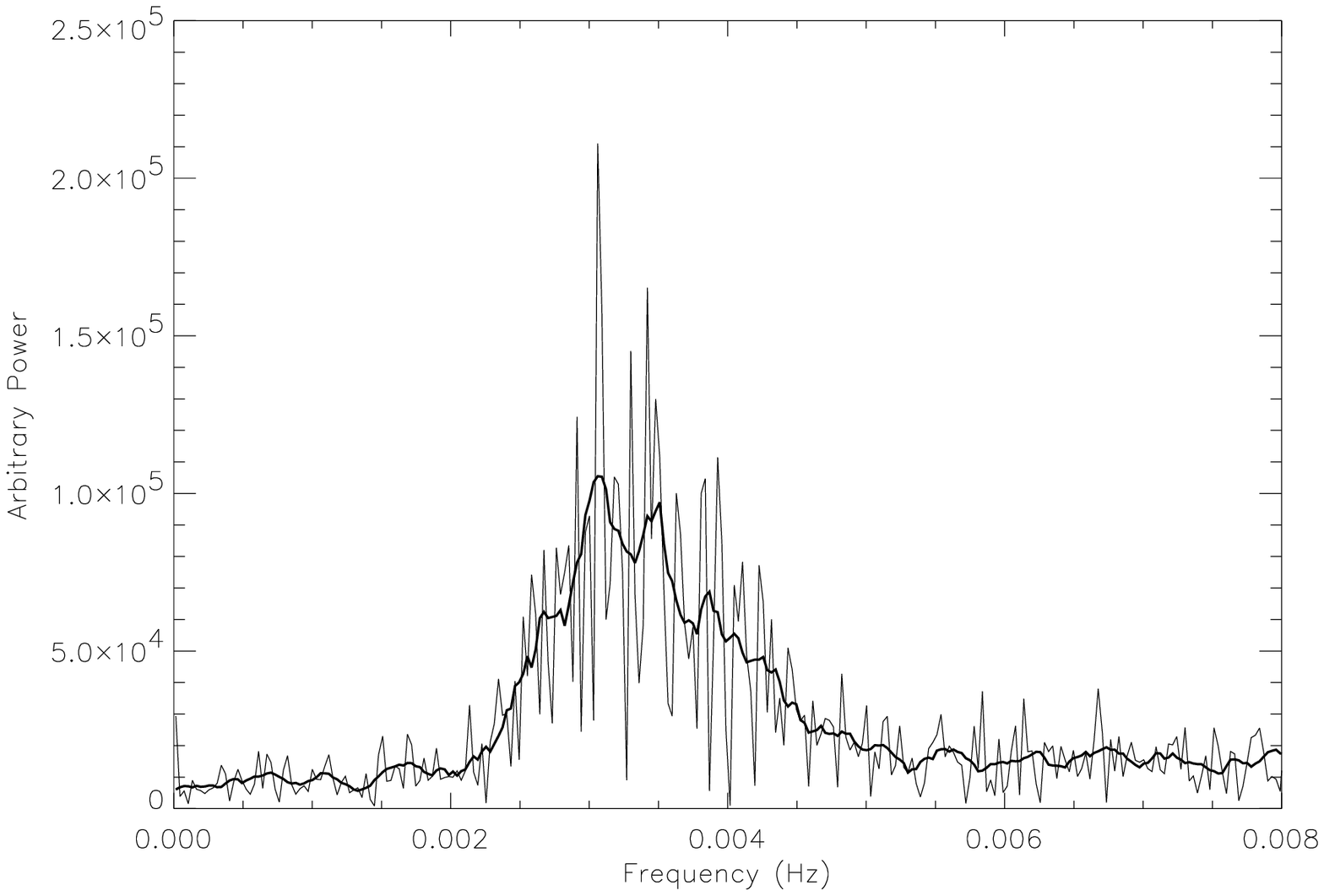}
\epsfxsize=2.75in\epsfbox{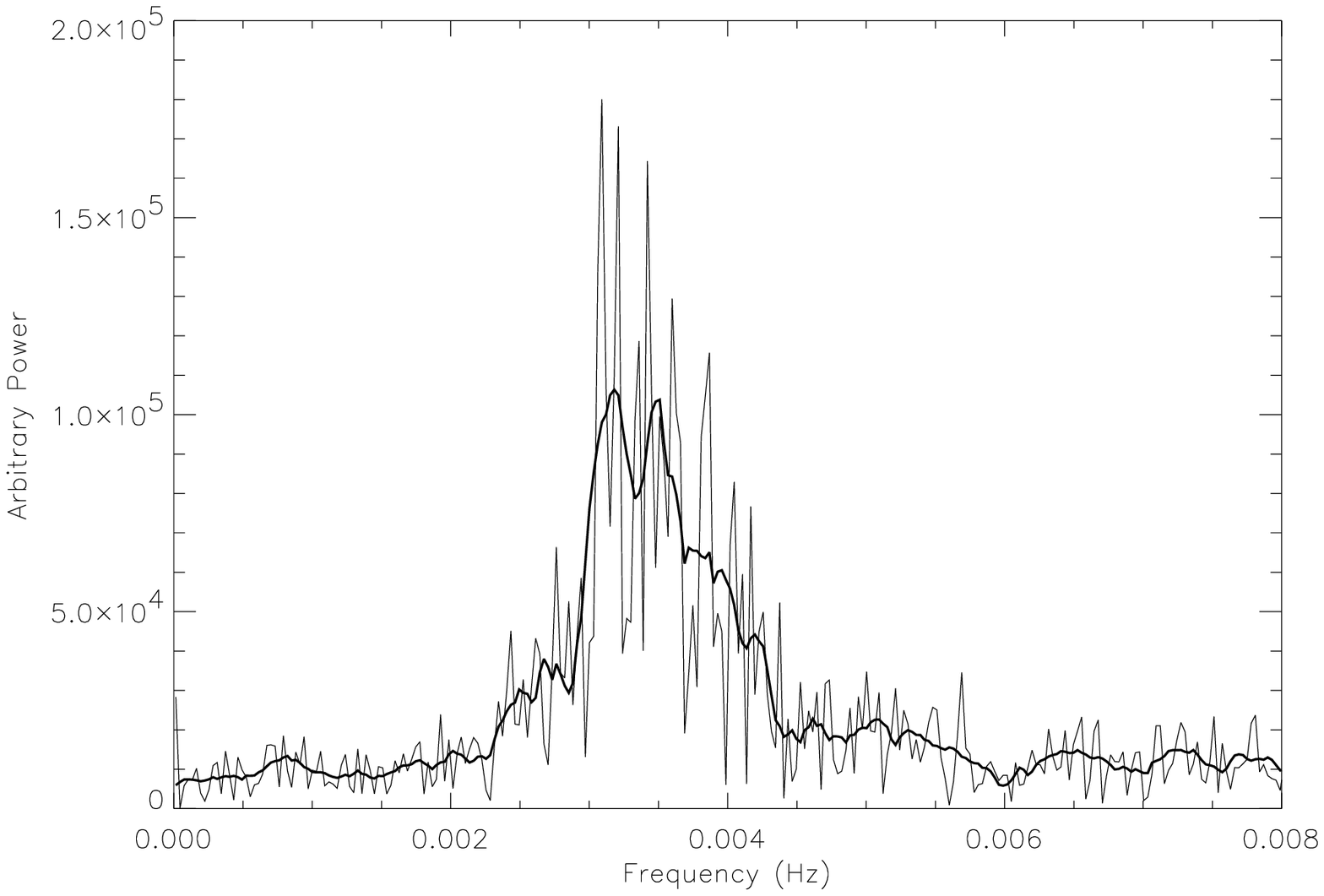}
\caption{Sample spectrum for 4 individual pixels{\it (thin solid lines)}. Shown in 
{\it heavy solid lines} are the fitted profiles estimated by S-G filter.} 
\end{center}
\end{figure}

\input epsf
\begin{figure}
\begin{center}
\leavevmode
\epsfxsize=5.0in\epsfbox{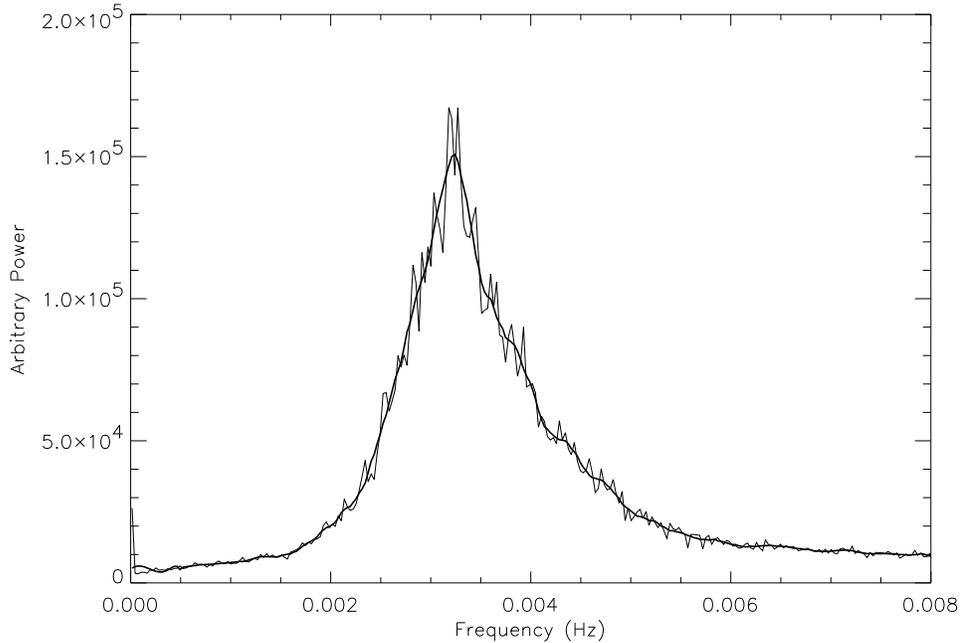}
\caption{Power spectrum averaged over 10 $\times$ 10 pixels {\it (thin solid line)}. Shown in 
{\it heavy solid line} is the fit estimated by S-G filter.} 
\end{center}
\end{figure}

\input epsf
\begin{figure}
\begin{center}
\leavevmode
\epsfxsize=5.0in\epsfbox{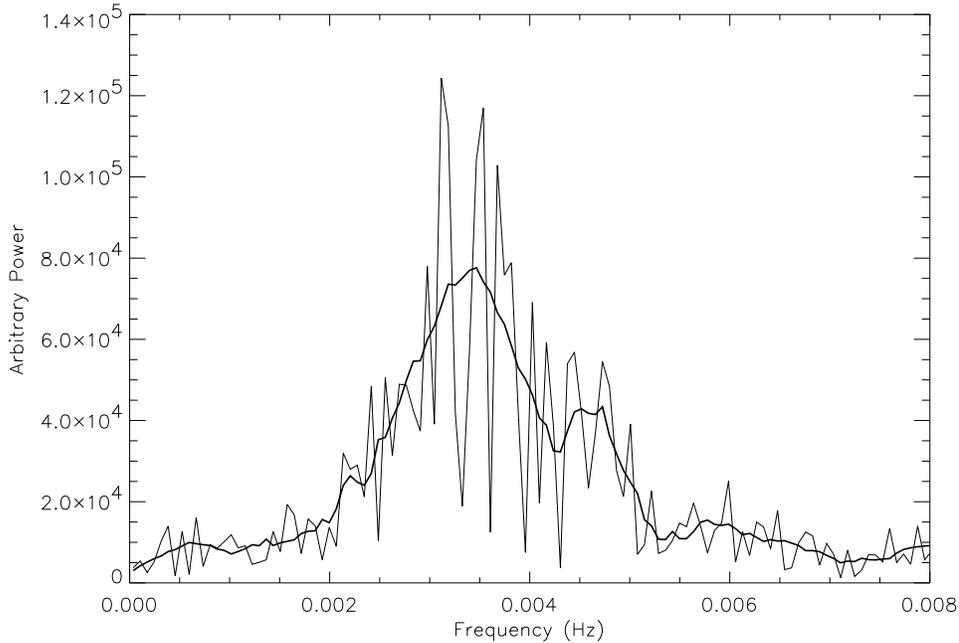}
\caption{Power spectrum for a single pixel{\it (thin solid line)} for a time baseline of 4 hrs as 
compared to 560 minutes for the original data as shown in Figure~1. Shown in 
{\it heavy solid line} is the fit estimated by S-G filter.} 
\end{center}
\end{figure}

\input epsf
\begin{figure}
\begin{center}
\leavevmode
\epsfxsize=5.0in\epsfbox{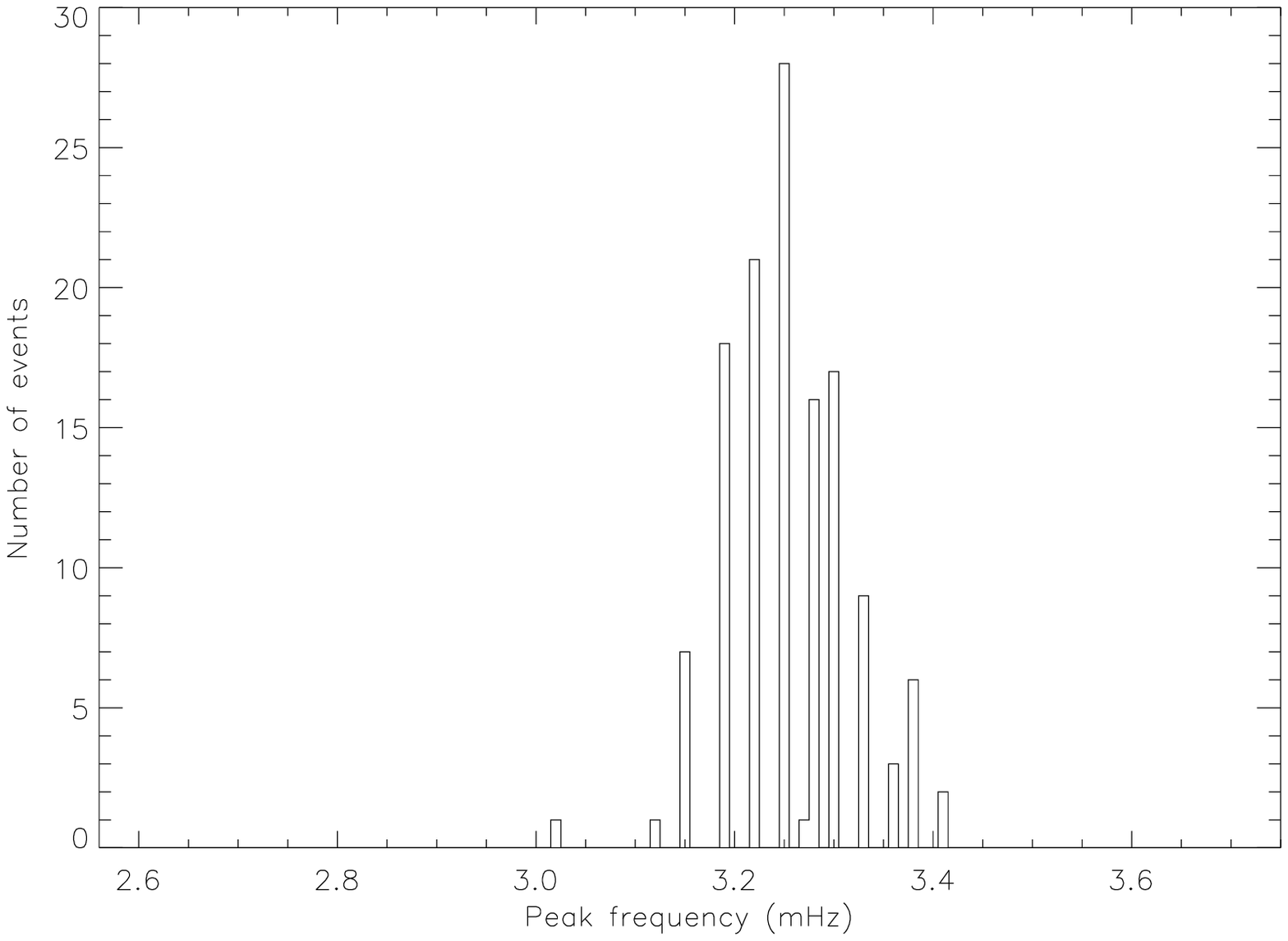}\\
\leavevmode
\epsfxsize=5.0in \epsfbox{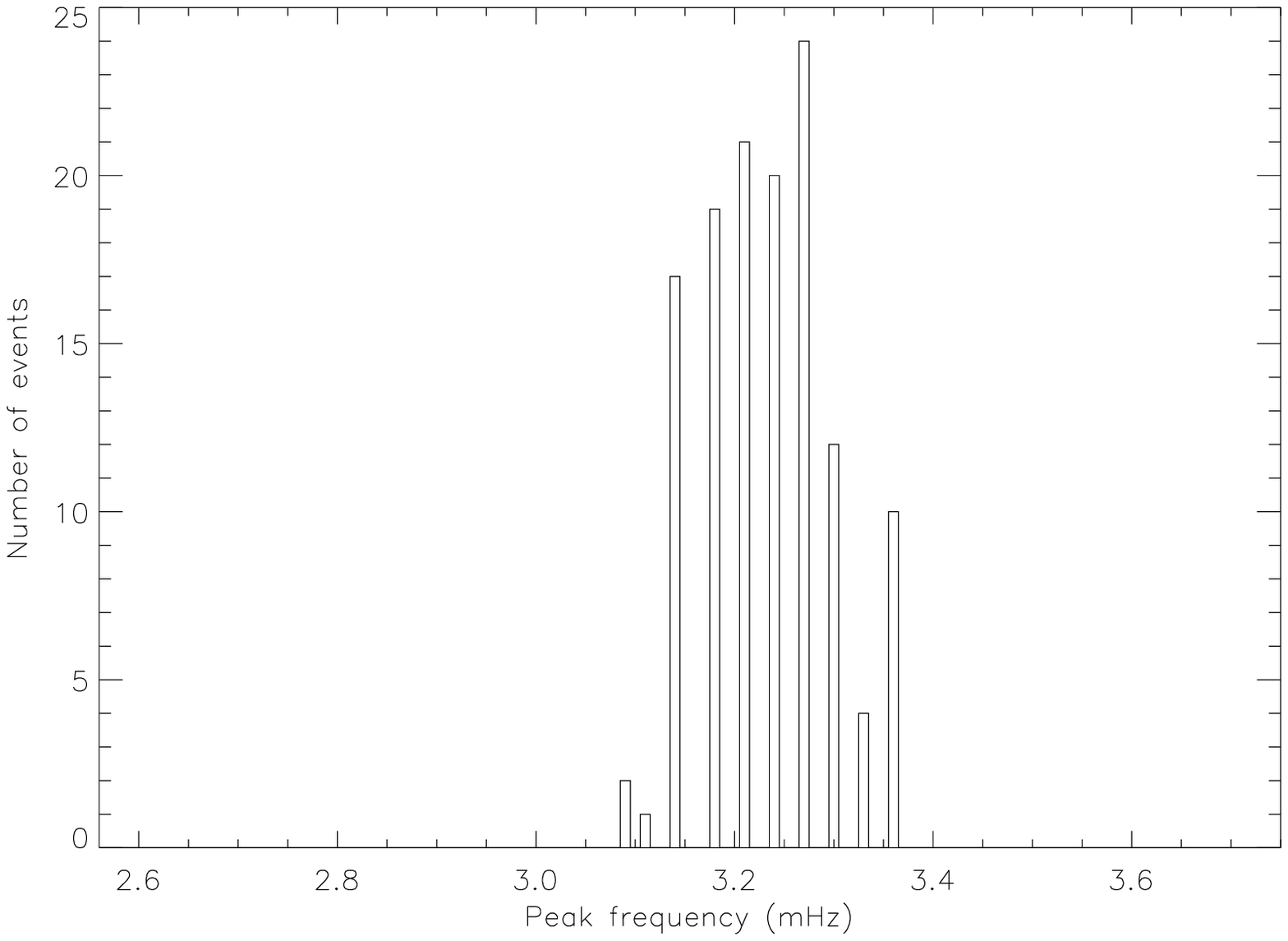}
\caption{Histograms of the distribution of the peak frequencies (for the 130 grids 
of the size 10 $\times$ 10 pixels) obtained 
for 11 May, 1997 {\it (top panel)} and 12 May, 1997 {\it (bottom panel)}.} 
\end{center}
\end{figure}

\input epsf
\begin{figure}
\begin{center}
\leavevmode
\epsfxsize=5.0in\epsfbox{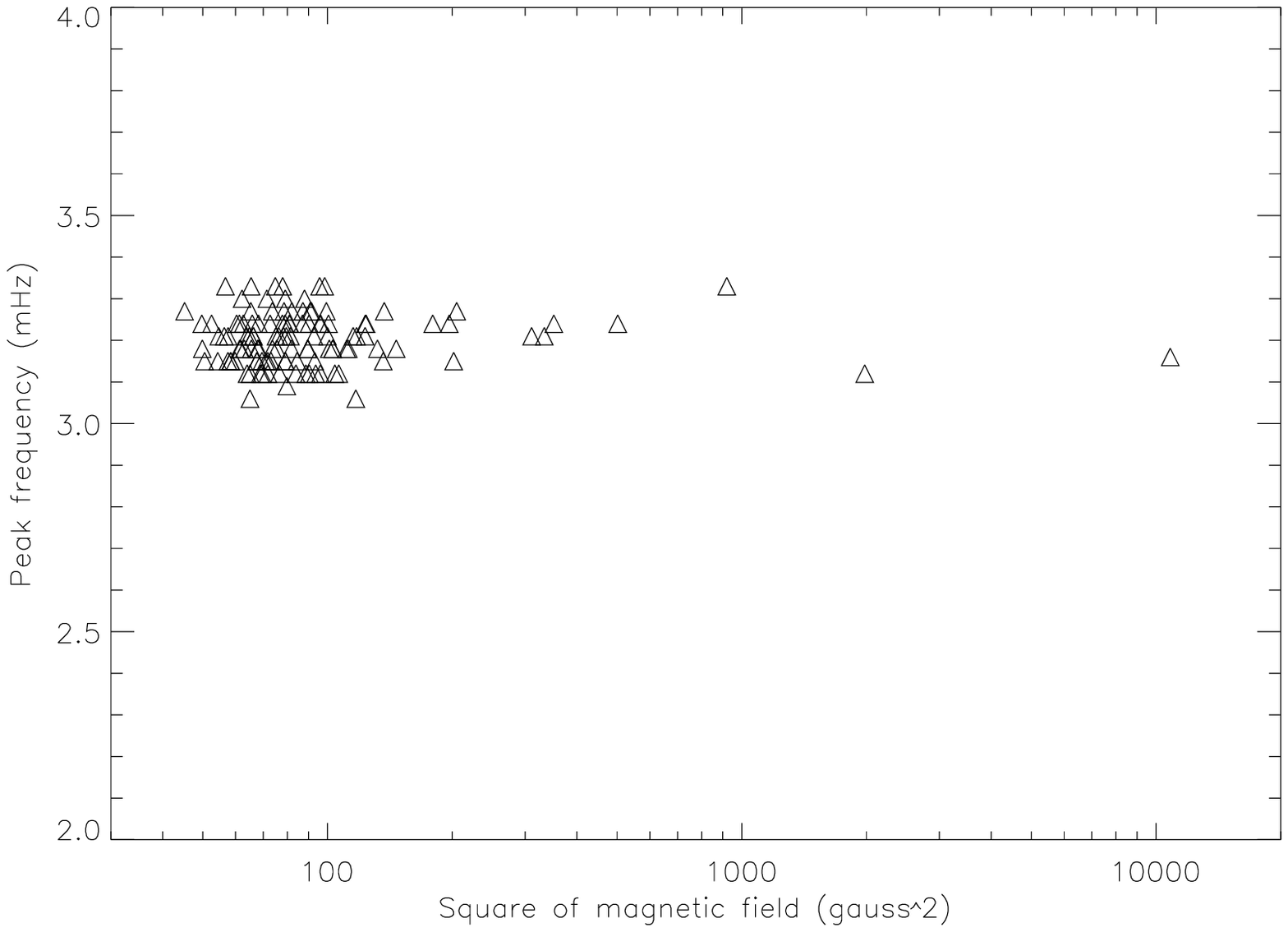}\\
\leavevmode
\epsfxsize=5.0in \epsfbox{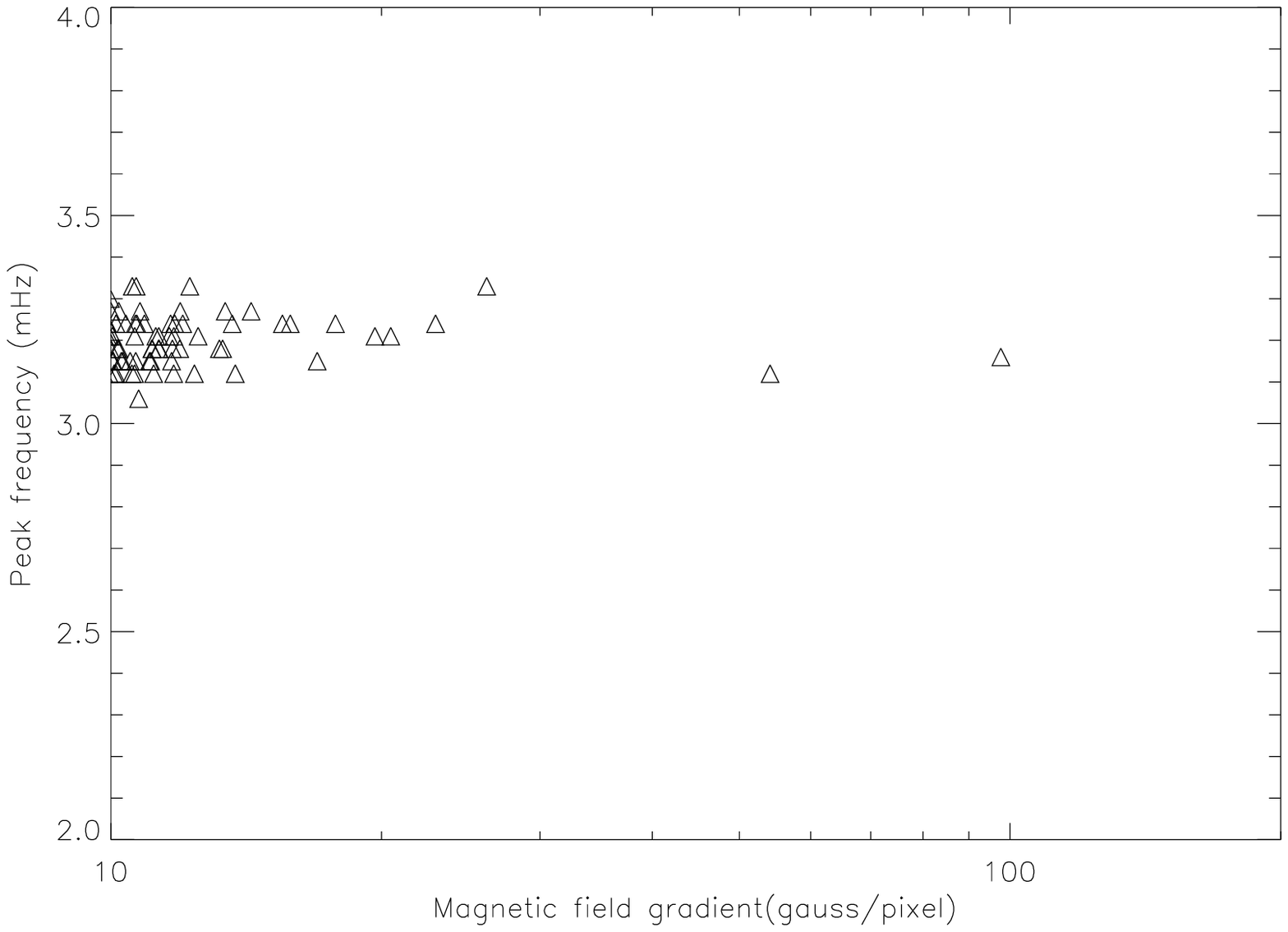}
\caption{Scatter plot of the peak frequencies (for the 130 grids of the size 10 $\times$ 10 pixels) 
against $B^2$ {\it (top panel)} and ${|{\nabla B}|}$ {\it (bottom panel)} averaged over 
the 100 pixels for 12 May, 1997.} 
\end{center}
\end{figure}

\input epsf
\begin{figure}
\begin{center}
\leavevmode
\epsfxsize=5.0in\epsfbox{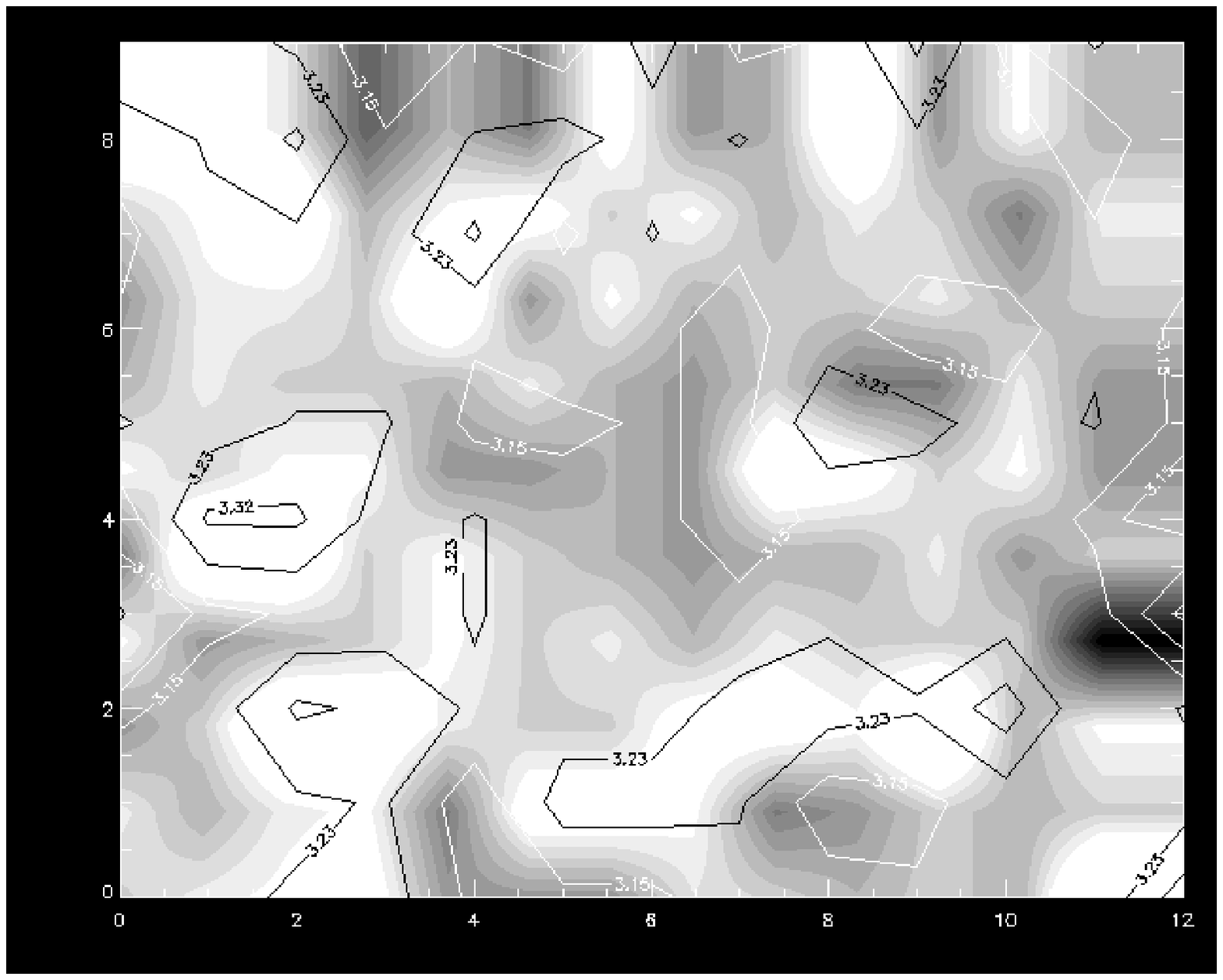}\\
\leavevmode
\epsfxsize=5.0in \epsfbox{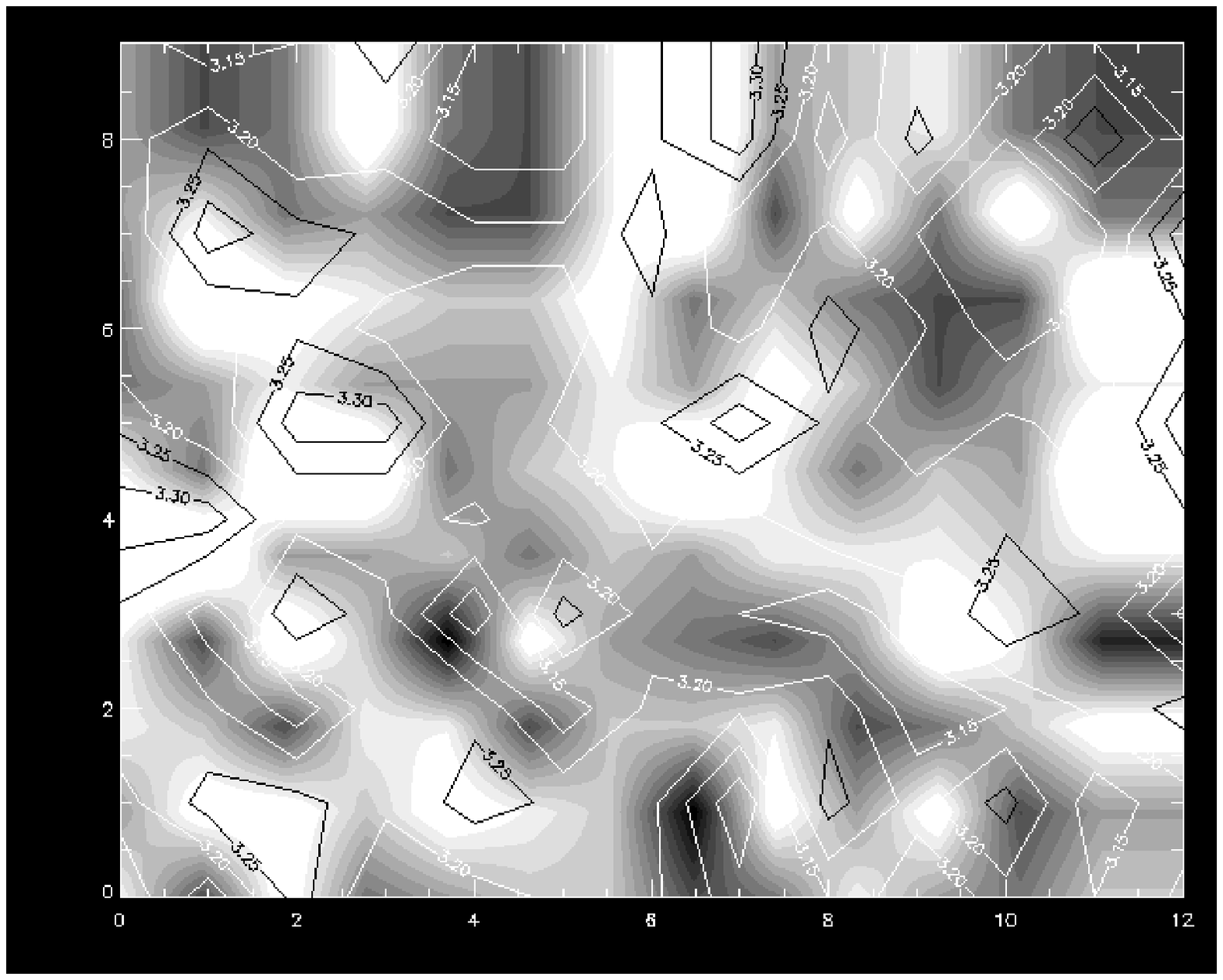}
\caption{Map of the peak frequencies (for the 130 grids of the size 10 $\times$ 10 
pixels) for 11 May, 1997 {\it (top panel)} and 12 May, 1997 {\it (bottom panel)}. The 
contours are overlaid on a grey scale image of the 2-D function depicting the variation 
of the peak frequency with latitude and longitude. The abscissa denotes 
50$^{\circ}$ E to 50$^{\circ}$ W and the ordinate 50$^{\circ}$ S to 50$^{\circ}$ N on the solar disk.} 
\end{center}
\end{figure}

\input epsf
\begin{figure}
\begin{center}
\leavevmode
\epsfxsize=5.0in\epsfbox{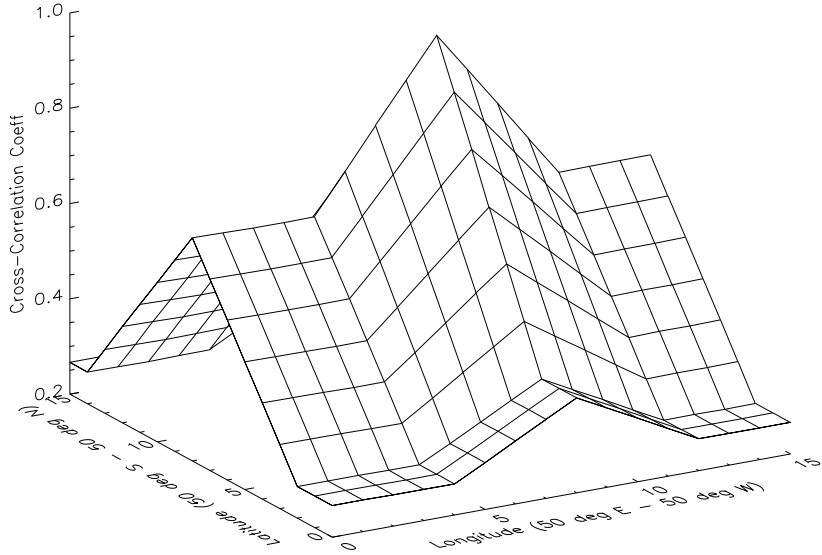}
\caption{Cross-correlationgram of the maps of peak frequencies for 11 and 12 May, 1997.} 
\end{center}
\end{figure}

\input epsf
\begin{figure}
\begin{center}
\leavevmode
\epsfxsize=5.0in\epsfbox{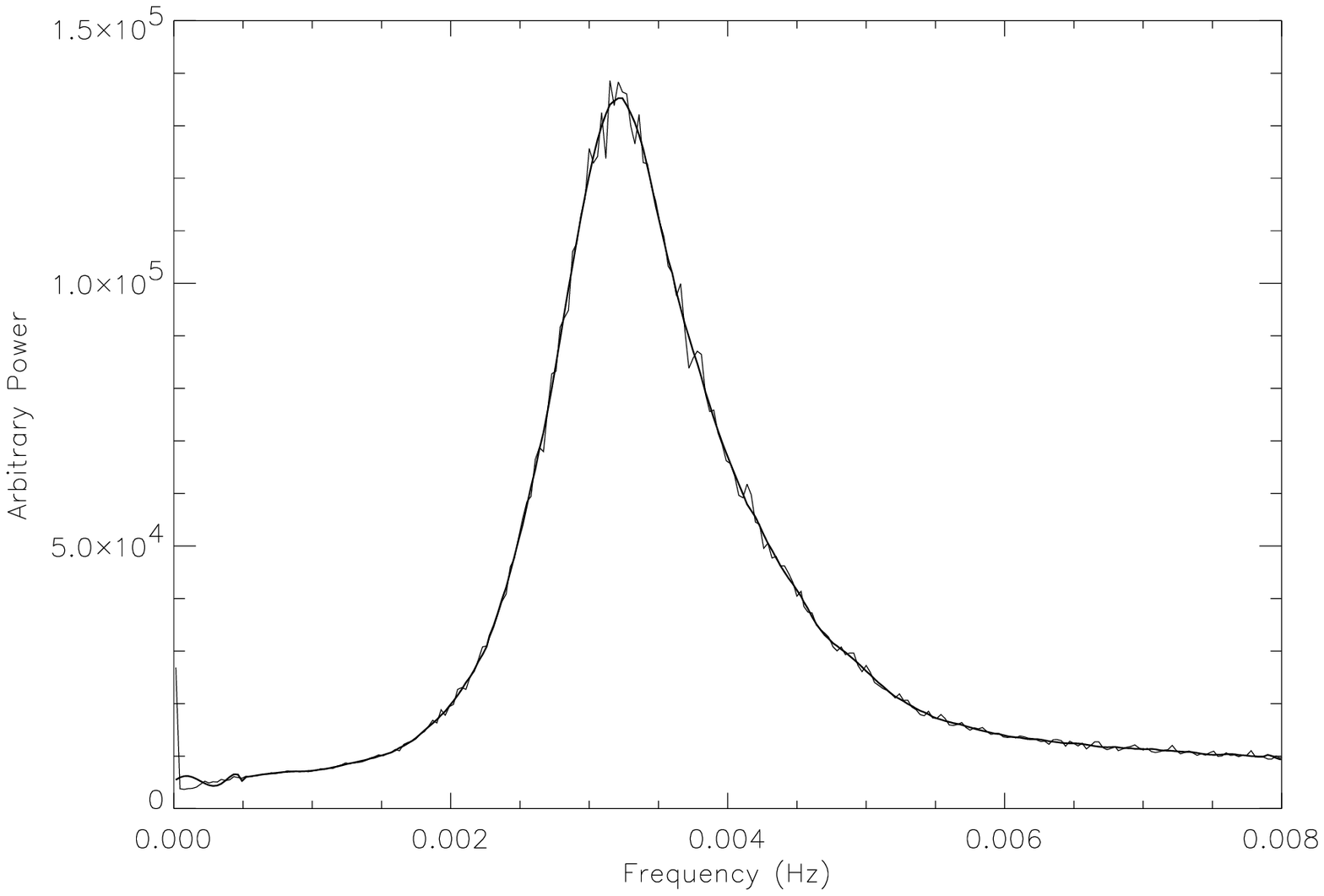}
\caption{Power spectrum averaged over 1300 pixels {\it (thin solid line)}. 
Shown in {\it thick solid line} is the fit estimated by S-G filter.} 
\end{center}
\end{figure}

\section{Results}
Figure~1 shows a sample spectrum for 4 individual pixels. It is seen
that the peak of the spectrum appears at different frequency positions
for the different spectra. We find that the {\it rms} variation of this peak
frequency is fairly independent of the number of points chosen for the
calculation of the {\it rms} value and is approximately 200 $\mu$Hz.
At first sight, it would seem that the acoustic spectrum has a high spatial variability. 
However, the solar spectrum is most likely to be powered by
a stochastic excitation process (Anderson, Duvall, and Jefferies, 1990). The
power spectrum of such a stochastically excited oscillation would be highly
variable, with peaks randomly appearing at various frequency positions due to
the superposition of a large number of independent natural oscillations with
random phases. If we assume that the power spectrum of each individual pixel
is an independent realization of this random process, then the sum of several
such spectra will produce more stable behaviour. We therefore summed 100 such
spectra from a 10 $\times$ 10 pixels patch to obtain an average spectrum.
In Figure~2, we have shown such an average power spectrum as well as the S-G fit 
applied to it for finding the peak frequency. It is seen that the uncertainty in
the position of the peak ($\sim$ 70 $\mu$Hz), as determined by the S-G fit, does 
not improve
with the averaging over 100 spectra. On the other hand, the uncertainty in fit 
almost doubles (146 $\mu$Hz) when we use a low resolution spectrum obtained 
from a 4 hrs sequence of the same data (Figure~3).
Coming back to the 560 minutes time series, the aforesaid S-G
algorithm is used to determine the peak of the spectrum 
for a raster of 130 grid points (each point providing an average spectrum 
for 10 $\times$ 10 pixels) spanning a rectangle of 
10 latitudes (from 50$^{\circ}$ S to 50$^{\circ}$ N) by 
13 longitudes (from 50$^{\circ}$ E to 50$^{\circ}$ W) on 
the Sun. It is to be noted that the size of the GONG Dopplergrams is different 
along the N-S and W-E directions because of the rectangular CCD pixels.
 
The histograms of the distribution of the peak frequencies obtained for
11 and 12 May, 1997 are shown in Figure~4. There is a
FWHM spread of 124 $\mu$Hz and 170 $\mu$Hz for 11 and 12 May respectively. 
The scatter
plots of these frequencies against $B^2$ and ${|{\nabla B}|}$ averaged over
10 $\times$ 10 pixels are shown in Figure~5. These plots show that the
magnetic effects, if any, are smaller than the spread in the frequencies. 
Figure~6 shows the ``map'' of the peak frequencies on 11 and 12 May respectively.
The cross-correlation of these two maps (Figure~7) shows no lag
corresponding to solar rotation, ruling out the existence of any ``feature'' with
lifetime greater than a day. The deviation in peak frequency as outlined by the 
contoured features in Figure~6 is not significantly greater than the half-width of the
histograms (Figure~4). These ``features'' therefore do not merit more attention.

We now look for variation in the peak frequency on very large scales. In particular,
we look for latitudinal variations. For this, we determine the average power spectrum of
a band consisting of 10 pixels in latitude and 130 pixels in longitude. 
Figure~8 shows one such spectrum. A set of 81 such spectra were generated by box-car
averages centred on a given latitude and spanning 10 pixels of latitude. We have
thus 40 spectra in N latitudes and 40 in S latitudes with 1 at the equator.
The mean of all the 40 peak frequencies at the N latitudes is 3.290 $\pm$0.020 mHz
while the mean of all the S latitude peak frequencies is 3.315 $\pm$0.015 mHz. The
difference in peak frequencies is 25 $\mu$Hz, which is barely greater than the {\it rms}
spread. It remains to be seen whether data with better spectral resolution will
be able to detect any N-S difference in the peak frequency. 

\section{Summary and Discussions}
Motivated by the various examples of spatial variability in the power of the
acoustic spectrum, we attempted to look for spatial variability in other parameters
of the spectrum,
which would be relatively less affected by the viewing angle. This concern about viewing
angle, arises out of the need to look for latitudinal variation in the acoustic
spectrum. The peak frequency in the acoustic spectrum seemed to be a good
candidate. However, the precision of the determination of this peak frequency on a 
spatial scale
of a single pixel (8 arc seconds for the GONG data) is limited by the stochastic 
variations in the power spectrum presumably caused by the
stochastic nature of the excitation process. Averaging over a large number of
spectra (100 spectra from a 10 pixel $\times$ 10 pixel area) produced more stable
spectra. The peak frequencies of 130 such locations were found to be distributed
with a FWHM
of about 130 $\mu$Hz. A map of the spatial variation of this peak frequency did not
show any strong feature with statistically significant deviation from the mean of
the distribution. Likewise, the scatter in the peak frequencies masked the
detection of magnetic field induced changes in the peak frequency. On a much larger
scale, the N latitudes showed a slightly lower
value of the peak frequency as compared to the S latitudes, although the
difference (25 $\mu$Hz) is barely larger than the {\it rms} spread (20 $\mu$Hz).

The N-S difference in the peak frequency needs to be examined with data having
higher spectral resolution. Similarly, the magnetic effects could be better
discerned at higher spectral resolution, as was indicated in MDI data (Lindsey,
2001, personal communication). The present work has shown that the spectra of
individual pixels can be treated as statistically independent realizations
of the stochastically varying acoustic spectrum. In the same way, statistically
independent realizations can be generated by using different initial times
of the time series of a single pixel. Averaging over many such spectra can produce
a stable spectrum for a single pixel. However, the detection of small scale
spatial variability in the peak frequency obtained from such ``temporally''
averaged spectra of individual pixels would nevertheless require better
spectral resolution. While the present work has shown the limits to which single 
station data can be pushed, a more significant ``acoustic map'' would perhaps be 
obtained from data having higher temporal resolution. 

\acknowledgements
This work utilizes data obtained by the Global Oscillation Network
Group (GONG) project, managed by the National Solar Observatory, a
Division of the National Optical Astronomy Observatories, which is
operated by AURA, Inc. under a cooperative agreement with the National
Science Foundation. The data were acquired by instrument operated by the 
Udaipur Solar Observatory. We are very much thankful 
to C. Lindsey, the referee, for his extensive comments, which helped 
in improving the manuscript significantly.

\end{article}
\end{document}